\documentclass[prl,showpacs,preprintnumbers,twocolumn]{revtex4}

\usepackage{amssymb}
\usepackage{amsmath}
\usepackage{epsfig}
\usepackage{graphics}
\usepackage[dvips]{color}

\providecommand{\unit}[1]{\,\mbox{#1}}

\begin{document}

\title{Hysteretic resistance spikes in quantum Hall ferromagnets
without domains}

\author{Henrique J. P. Freire}

\author{J. CarlosEgues} 

\affiliation{ Departamento de F\'{\i}sica e Inform\'{a}tica, Instituto
de F\'{\i}sica de S\~{a}o Carlos, Universidade de S\~{a}o Paulo,
13560-970 S\~{a}o Carlos, S\~{a}o Paulo, Brazil}

\date{December 17, 2004}

\begin{abstract}
We use spin-density-functional theory to study recently reported
hysteretic magnetoresistance $\rho _{xx}$ spikes in Mn-based 2D
electron gases [Jaroszy\'{n}ski \textit{et al.} Phys. Rev. Lett.
\textbf{89}, 266802 (2002)]. We find hysteresis loops in our
calculated Landau fan diagrams and total energies signaling
quantum-Hall-ferromagnet phase transitions. Spin-dependent
exchange-correlation effects are crucial to stabilize the relevant
magnetic phases arising from \textit{distinct} symmetry-broken
excited- and ground-state solutions of the Kohn-Sham equations.
Besides hysteretic spikes in $\rho _{xx}$, we predict
\textit{hysteretic dips} in the Hall resistance $\rho _{xy}$. Our
theory, \textit{without} domain walls, satisfactorily explains the
recent data.
\end{abstract}

\preprint{Journal reference: Phys.~Rev.~Lett.~\textbf{99}, 026801 (2007)}%


\maketitle

Two-dimensional electron gases (2DEGs) under strong magnetic fields
exhibit fascinating physical phenomena; the mostly widely known of
these being the integer and fractional quantum Hall effects
\cite{sarma-pinc}. Spontaneous magnetic order in quantum-Hall systems
is yet another interesting possibility. Quantum Hall ferromagnetism
arises from the interplay of the Zeeman, Coulomb, and thermal energies
within the macroscopically degenerate Landau levels of the 2DEG
\cite{quinn,disorder}. Landau level crossings \cite{fang-stiles} offer
a convenient means to probe symmetry-broken quantum-Hall ferromagnetic
transitions. At crossings, opposite-spin levels can benefit from
Coulomb exchange to form spin-ordered states at low temperatures
\cite{quinn}.

Many groups have investigated quantum Hall ferromagnetism in the
integer and fractional quantum-Hall regimes by inducing Landau level
crossings via tilted magnetic fields, density and level tuning via
gate electrodes, hydrostatic pressure \cite{hysteresis}, and the
\textit{s-d} exchange-induced level bowing in Mn-based 2DEGs
\cite{knobel,jaros}. These studies \cite{hysteresis,knobel,jaros}
find ubiquitous ``anomalous'' peaks and hysteretic spikes in
Shubnikov-de-Haas measurements of the magnetoresistance $\rho_{xx}$.
The hysteretic behavior here means that the spikes appear at
distinct magnetic fields as the field is swept up and down. In the
fractional quantum-Hall regime recent experiments show that these
features follow from the hyperfine coupling between electrons and
nuclei \cite{stern}. In the integer quantum Hall regime, the
hysteretic spikes have been suggested to arise from charge transport
along long domain-wall loops acting as one-dimensional
(``percolating'') channels in the 2DEG near an Ising-like
quantum-Hall ferromagnet transition \cite{jun-mac}. Though
appealing, this description is largely qualitative: no
magnetotransport quantities (e.g., $\rho_{xx}$) have been calculated
so far accounting for domain walls.
\begin{figure}[t]
\begin{center}
\includegraphics[width=8.6cm,keepaspectratio=true]{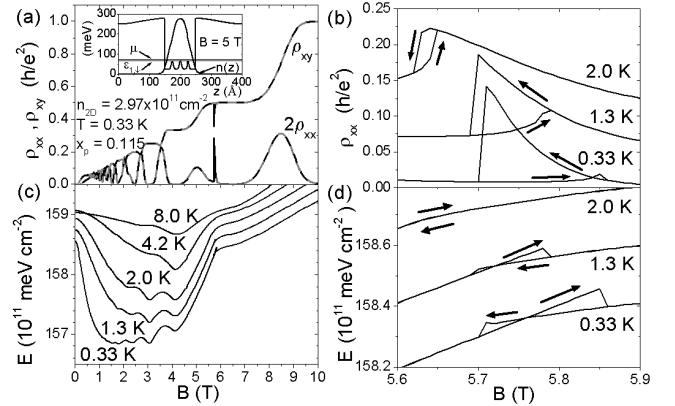}
\end{center}
\caption{Hysteretic magnetoresistance $\rho_{xx}$, Hall resistance
$\rho_{xy}$, and total energy $E$, as a function of $B$ for a
quantum well [inset in (a)]. For the up$ \rightleftharpoons$down
\textit{B} sweeps, hysteretic spikes and dips appear in
$\protect\rho_{xx}$ and $\protect\rho_{xy}$, respectively, at
$B\sim5.8$ T (a). For increasing temperatures the hysteretic loops
become less pronounced ($\protect\rho_{xy}$ not shown) and disappear
above a critical temperature $T_c\sim2.1$ K (b). The total energy
$E$ (c) shows hysteretic loops [see blow-up of the loops in (d)]
because the KS equations have two distinct solutions (ground and
excited states), e.g., for $B_c^d <B<B_c^u$. These are quantum-Hall
ferromagnet phases with distinct spin polarizations (Fig.~2). The
curves in (b) are displaced vertically for clarity.}
\label{figResCdTe}
\end{figure}

\textit{Altenate picture for the hysteretic spikes.} In this Letter
we apply the Spin Density Functional Theory (SDFT) \cite{dft}
implemented via the Kohn-Sham (KS) scheme in a Local Spin Density
Approximation (LSDA) to determine the electronic structure of a
2DEG, which we then use in a linear response model
\cite{ando-uemura} to explicitly calculate $\rho_{xx}$ and
$\rho_{xy}$. For concreteness, we focus on the experiment of
Jaroszy\'{n}ski \textit{et al.} \cite{jaros} in Mn-based 2DEGs
\cite{peaks-nms}. Interestingly, we find hysteretic behavior in
$\rho_{xx}$ and $\rho_{xy}$, Figs.~1(a)-(b), \textit{without} taking
into account domain walls.  This behavior follows from our system
having two self-consistent KS solutions -- for a \textit{same} set
of parameters -- with distinct total energies as shown in
Figs.~1(c)-(d), see e.g.~range $B_c^d<B<B_c^u$. As we discuss later
on, these solutions correspond to ground and excited states
describing distinct quantum Hall ferromagnetic phases of our
interacting 2DEG, and, most importantly, provide a concrete example
for a theorem of Perdew and Levy on the existence of excited states
from ground-state density functionals \cite{perdew-levy}. These
phases comprise differing sets of conducting states contributing to
the magnetotransport and hence have distinctive $\rho_{xx}$'s and
$\rho_{xy}$'s. The hysteresis then arises because the 2DEG can go
through a different sequence of magnetic phases (i.e., the system
can become trapped into distinct local minima) as the $B$ field is
swept up and then down, e.g., phases I to II in the up sweep at
$B\simeq B_c^u$ and phases III to IV in the down sweep at $B\simeq
B_c^d \neq B_c^u$, Fig.~1(d). We predict \textit{hysteretic dips}
(and peaks \cite{peaks-nms}) in $\rho_{xy}$ \cite{jaros-priv}
[Fig.~1(a)] and that the spikes shift to opposite directions
(Fig.~3) in samples with positive and negative $g$ factors, for
increasing tilt angles of $B$.

\textit{Mn-based system.} We consider a CdTe quantum well
between Cd$_{0.8}$Mg$_{0.2}$Te barriers, with three evenly spaced Cd$_{1-x_p} $Mn$_{x_p}$Te monolayers (``Mn barriers'') in the well
region \cite{jaros}, Fig. 1(a) (inset); $x_p$ is the planar
concentration of Mn. Adjacent to the barriers lie two symmetric
n-doped regions \cite{exp-doping}. In an external field $B$, the
\textit{s-d} exchange interaction between the electrons in the well
(2DEG) and those of the localized \textit{d} orbitals of the Mn
gives rise to a spin-dependent electron potential
\begin{equation}
v_{s\text{-}d}^{\sigma_{z}}\left( z;B,T\right) =\frac{\sigma_{z}}{2}N_{0}\alpha
\bar{x}(z)\frac{5}{2}B_{5/2}\left[ \frac{5\mu_{B}B}{k_{B}\left(
T+T_{0}\right) }\right] ,  \label{eq1}
\end{equation}
where $N_0\alpha$ is the \textit{s-d} exchange constant, $B_{5/2}$ is the
spin-5/2 Brillouin function, $\bar x(z)$ and $T_0$ are the effective Mn
profile and temperature \cite{furdyna}, respectively, and $\sigma_z=\pm1$
(or $\uparrow,\downarrow$) denotes the electron spin components. The
structural confining potential of the well is assumed square $v_w(z)=v_0[\Theta(-L/2-z)+\Theta(L/2+z)]$ with depth $v_0$; $\Theta(z)$ is
the Heaviside function. The structural potential of the Mn barriers is $v_b(z)=v_1x(z)$; $v_1$ is the barrier height and $x(z)$ the nominal gaussian
Mn profile.

\textit{Kohn-Sham approach.} We use the SDFT \cite{dft} formulated
in the context of the effective-mass approximation of
semiconductors. Within the finite-temperature formulation of Mermin
\cite{mermin,gunn-lund}, we obtain the KS equations
\begin{equation}
\left[ -\frac{\hbar^2 }{2m}\frac{d }{d z^{2}}+v_\text{ef\/f}^{\sigma _{z}}\left(
z;[n_{\uparrow },n_{\downarrow }]\right) \right] \chi _{i}^{\sigma
_{z}}\left( z\right) =\varepsilon _{i}^{\sigma _{z}}\chi _{i}^{\sigma
_{z}}\left( z\right),  \label{eq2}
\end{equation}
where $m$ is the effective mass, $i=1,2\ldots $ the band index, and
$v_\text{ef\/f}^{\sigma_{z}}(z;[n_\uparrow,n_\downarrow])$ the effective
single-particle potential
\begin{equation}\begin{split}
v_\text{ef\/f}^{\sigma_z}(z;[n_\uparrow,n_\downarrow])
   = & \, v_{s}(z) 
   + v_{s\text{-}d}^{\sigma_z}(z) 
   + v_{h}(z;[n]) \\
   & + v_{xc}^{\sigma _{z}}(z;[n_\uparrow,n_\downarrow ]) \, .  
\label{eq3}
\end{split}
\end{equation}
In Eq. (\ref{eq3}) $v_{s}(z)=v_{w}(z)+v_{b}(z)$, $v_{h}(z;[n])$ is the
Hartree potential, calculated by solving Poisson's equation, and
$v_{xc}^{\sigma _{z}}(z;[n_{\uparrow },n_{\downarrow }])$ is the
\textit{spin-dependent} exchange-correlation (XC) potential
\cite{vosko80}. The motion in the \textit{xy} plane is quantized into
Landau levels with energies $\varepsilon _{n}=(n+1/2)\hbar \omega
_{c}$, $n=0,1,2,\ldots $ and $\omega _{c}=eB/m$ \cite{LL}. The total
wavefunction is $\psi _{i,n,k_{y}}^{\sigma
_{z}}(x,y,z)=\frac{1}{\sqrt{L_{y}}}\exp(\mathrm{i}k_{y}y)\varphi _{n}(x)\chi
_{i}^{\sigma _{z}}\left( z\right)$, where $\varphi _{n}(x)$ is the
\textit{n}-th harmonic oscillator eigenfunction centered at
$x_{0}=-\hbar k_{y}/m\omega _{c}$ and $k_{y}$ is the electron wave
number along the $y$ axis; $L_{y}$ is a normalizing length. This
decoupling of the $z$ and \textit{xy} motions follows from the
uniformness of the total electron density within the \textit{xy}
plane: because each electron can be anywhere within the plane, we use
the average total electron density $n(z)=\sum_{i,n,k_{y},\sigma
_{z}}f_{i,n}^{\sigma _{z}} \frac{1}{L_xL_y}\int \int |
\psi_{i,n,k_{y}}^{\sigma _{z}}| ^{2}dxdy=\sum_{i,n,k_y,\sigma
_{z}}|\chi _{i}^{\sigma _{z}}(z)|^{2}f_{i,n}^{\sigma _{z}}$ in
Poisson's equation, instead of $ n(x,y,z)$ [here $f_{i,n}^{\sigma
_{z}}$ is the Fermi function and $L_x$ is a normalizing length]. This
procedure makes the 2DEG uniform thus rendering a separable KS set.

We assume that the KS eigenvalues
\begin{equation}
\varepsilon _{i,n}^{\sigma _{z}}=\varepsilon _{i}^{\sigma _{z}}(B)+\left( n+
\frac{1}{2}\right) \hbar \omega _{c}+\frac{\sigma _{z}}{2}g\mu _{B}B,
\label{eq4}
\end{equation}
where $g\mu_B B \sigma_z/2$ is the ordinary Zeeman term ($g$:
effective Land\'e factor), describe the actual electronic structure
of our 2DEG. This assumption is, in principle, unjustified within
DFT: the individual KS orbitals represent states of a fictitious
non-interacting electron gas in an effective potential, Eqs.
(\ref{eq2}) and (\ref{eq3}). With this assumption, however, we
satisfactorily explain observed hysteretic phenomena in 2DEGs
\cite{jaros,peaks-nms}.

\textit{System parameters.} In our simulations we use (see Ref.
\cite{jaros} ): $m/m_0=0.099$, dieletric constant
$\epsilon/\epsilon_0=10$, $g=-1.67$, $ N_0\alpha=220$ meV,
quantum-well width $L=100\unit{\text{\AA}}$, spacer width
$L_s=200\unit{\text{\AA}}$, $n_{2D}=2.97\times 10^{11}$ cm$^{-2}$,
number of Mn monolayers $N_b=3$, $ x_p=0.115$,
$v_0=248.1\unit{meV}$, $v_1=1183.5\unit{meV}$, $T_0=0.47$ K, and
assume a diffusion length $\ell\sim 4.67 \unit{\text{\AA}}$ for the
Mn profile.

Figures 1 and 2 show our theoretical results for $\rho_{xx}$,
$\rho_{xy}$ 1(a)-(b), total energy $E$ 1(c)-(d), Landau level fan
diagram 2(a), spin-resolved electron densities
$n_{2D}^{\uparrow,\downarrow}$ 2(b), and spin-polarization $\zeta$
2(c). Remarkably, all of these quantities show abrupt hysteretic
changes near 5.8 T as the $B$ field is swept up and down. The
discontinuous nature of these features follows from the 2DEG
undergoing (quantum) phase transitions in which its degree of spin
polarization $\zeta$ suddenly changes, 2(b)-(c), e.g., in the down
sweep the 2DEG becomes highly spin-polarized near 5.8 T [2(c)]: a
quantum Hall ferromagnetic phase transition takes place. Our
calculated fan diagram 2(a) and total energy 1(d) corroborate this
scenario: the opposite-spin levels $\varepsilon_{1,0}^{\uparrow}$,
$\varepsilon_{1,1}^{\downarrow}$ suddenly cross near $\mu$ thus
lowering the total energy  $E$ [see III to IV discontinuity in
1(d)]. This is similar to the Giuliani-Quinn instability where the
$1^{\rm st}$-order transition is due to the gain of exchange energy
in the ferromagnetic state \cite{quinn}. Our calculation also
includes \textit{correlation} which somewhat reduces the exchange
effects. The role of the \textit{s-d} exchange [Eq. (\ref{eq1})] in
our system is to cause ``level bowing'' ($B<2$ T) thus inducing
opposite-spin Landau level crossings, see e.g. crossings for $B<4$ T
in 2(a). When crossings occur near $\mu$, spin-dependent XC effects
may induce phase transitions (e.g. at 5.8 T); we do not find phase
transitions in a Hartree calculation.  The phases here are those of
an Ising-like itinerant 2D ferromagnet with Pauli magnetization $
(n_\uparrow-n_\downarrow)\mu_B = n_{2D}\zeta \mu_B$. Next we show
how hysteresis arises in our SDFT calculation.

\begin{figure}[t]
\begin{center}
\includegraphics[width=7.4cm,keepaspectratio=true]{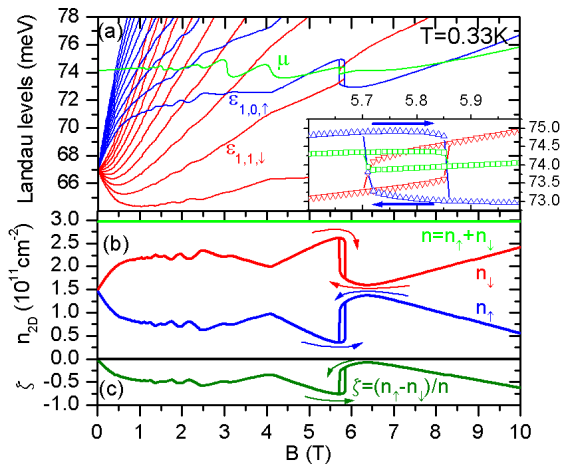}
\end{center}
\caption{Landau-level fan diagram [Eq. \protect\ref{eq4}] (a),
spin-dependent electron densities $n_{2D}^{\uparrow,\downarrow}$ (b), and
spin polarization $\protect\zeta$ (c) of our 2DEG [Fig. 1(a)]. In (a) we
plot the center of each Landau level with a phenomenological gaussian broadening $
\Gamma=0.36 \protect\sqrt{B} \unit{meVT}^{-1/2}$
\protect\cite{ando-uemura}. The \textit{s-d} induced non-linear
behavior of the energy spectrum ($B<2$ T) allows for non-trivial
Landau level crossings near the chemical potential $\protect\mu$
(curve at $\sim 74 \unit{meV}$). Inset (a): blow-up of the
hysteretic crossings between the states with $n=1$ and spin down ($\protect
\varepsilon_{1,1}^\downarrow$) and that with $n=0$ and spin up ($\protect
\varepsilon_{1,0}^\uparrow$) at $B\sim 5.8\unit{T}$ for the up$
\rightleftharpoons$down \textit{B} sweeps. The discontinuous change
in the spin-polarization $\protect\zeta$ signals a transition
between distinct quantum-Hall ferromagnet phases. We have adjusted
$x_p$ and $\Gamma$ so the
levels would cross at $\sim 5.8\unit{T}$ as in the experimental data \protect\cite{jaros}. }
\label{figResCdTeHist}
\end{figure}

\textit{Excited and ground Kohn-Sham states}. Using the
constrained-search definition of the ground state functional
$E_v[n]$ (defined on domain of densities constructed from any wave
function), Perdew and Levy  \cite{perdew-levy} have shown that
\noindent \textit{``...every extremum of $E_v[n]$ represents the
density $n_i(\textbf{r})$ and the energy $E_i$ of a stationary
state. The absolute minima represent the ground states, and the
extrema lying above the minimum represent a subset of the excited
states.''} \noindent These authors have also proved that some of the
self-consistent solutions of the KS equations extremize $E_v[n]$,
provided that these solutions obey ground-state Fermi statistics
(\textit{aufbau} principle) with a \textit{single} chemical
potential -- this is a necessary and sufficient condition. These
theorems have been generalized to SDFT \cite{perdew} and hence hold
in our system.

Our simulations show that for a certain window of magnetic field
[e.g., $B_c^d < B < B_c^u$ in Fig. 1(d)] the KS equations
(\ref{eq2}) have indeed two self-consistent solutions
\cite{self-consistency} with distinct total energies  and spin
polarizations \cite{gunn-lund} -- both satisfying the necessary and
sufficient condition above. At each $B$ in this range, one of these
two solutions is stable (true minimum, ground state) and the other
metastable (local minimum, excited state). In practice, these two
states (phases) are separated by an energy barrier \cite{barrier}
which may trap the system in the metastable states during the up$
\rightleftharpoons$down \textit{B} sweeps thus giving rise to
hysteretic loops, Figs. 1 and 2.

\textit{Magneto-transport.} We obtain the longitudinal and Hall
resistances from the conductivity tensor
($\mathbf{\rho}=\mathbf{\sigma}^{-1}$), calculated within the
self-consistent Born-approximation model of Ando and
Uemura \cite{ando-uemura}. For short-range scatterers, $\sigma_{xx} = \frac{
4e^2}{\hbar} \int\limits_{-\infty}^{\infty} \left( -\frac{\partial
f(\varepsilon)}{\partial \varepsilon} \right) \notag  \sum_{i,n,\sigma_z}
\left(n+\frac{1}{2}\right)\exp \left[ - \left( \frac{\varepsilon-
\varepsilon_{i,n}^{\sigma_z}} {\Gamma_{\mathrm{ext}}} \right)^2 \right]
\mathrm{d} \varepsilon $ and $\sigma_{xy} = e n_{2D}/B + \Delta\sigma_{xy}$;
$\Delta \sigma_{xy}$ is a small correction \cite{ando-uemura}. We model the
extended Landau states by a gaussian $g_{\mathrm{ext}}(\varepsilon)$ of
width $\Gamma_{{\mathrm{ext}}}=0.25$ meV \cite{prange}.

\textit{Hysteretic resistance spikes and dips.} The ordinary $\rho
_{xx}$ peaks in Fig.~1(a), e.g., at 2.5, 3.1, and 5 T, are due to
subsequent \textit{single} Landau levels crossing the chemical
potential $\mu$ and appear between plateaus in $\rho_{xy}$. Though
involving the crossing of \emph{two} opposite-spin levels at $\mu $,
the $\rho _{xx}$ spike at $B\sim 5.8 \unit{T}$ has the same physical
origin as the ordinary peaks: electrons in partially filled crossing
levels are susceptible to scattering (``dissipation'') which increases
$\rho_{xx}$. This spike, however, corresponds to a dip in a plateau
of $\rho_{xy}$: as the two Landau levels cross near $\mu$, the
number of conducting channels $n_{c}$ \textquotedblleft
fluctuates\textquotedblright\ thus making $\rho _{xy}$ dip ($\Delta
n_{c}>0$ ) [or peak ($\Delta n_{c}<0$)] toward an adjacent plateau
consistent with $n_{c}$.  Whether $\rho _{xy}$ dips or peaks and
whether the spikes and dips are stronger in the up or down sweeps
depend on the details of the crossings at $\mu $, Fig.~2(a) -- these
features are, however, hysteretic as $n_c$ differs in the up and down
$B$ sweeps. For increasing tilt angles $\theta $, Fig.~3, the
hysteretic spike shifts to lower fields while the ordinary
Shubnikov-de-Haas maxima do not. Figure 3 uniquely identifies ordinary
(single Landau levels crossing $\mu$) and \textquotedblleft anomalous
peaks\textquotedblright\ (two Landau levels crossing $\mu$) in
$\rho_{xx}$.

\begin{figure}[b]
\begin{center}
\includegraphics[width=7.5cm,keepaspectratio=true]{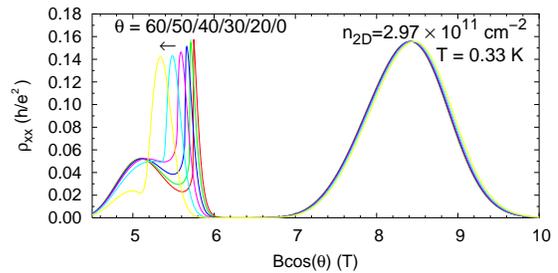}
\end{center}
\caption{Magnetoresistance $\protect\rho_{xx}$ vs
$B\cos(\protect\theta)$ for several tilt angles $\protect\theta$
between the $B$ field and the growth direction. Similarly to the
data in \protect\cite{jaros}, the spike shifts to lower fields as
$\protect\theta$ increases, while the ordinary Shubnikov-de-Haas
peaks do not.} \label{fig3}
\end{figure}

\textit{Hysteresis $\And$ critical temperature.} Figure 4(a) shows
the hysteresis in the $\rho_{xx}$ spike position $B_c$ as a function
of $T$: $B_c^{u}>B_c^{d}$ for the up $\rightleftharpoons$down $B$
sweeps. By plotting $\Delta B_c=B_c^{u}-B_c^{d}$ versus $T$ we can
extract a critical temperature $T_c$ above which $\Delta B_c=0$. We
find [inset in 4(a)] $T_c=2.1\unit{K}$ which is comparable to the
experimental value 1.3 K \cite{jaros}. The \textit{amplitude} of the
spike is also hysteretic, Fig. 4(b). Here, however, only the
up-sweep behavior agrees with the data \cite {jaros,exp-doping}.

\textit{Further comparison with experiments.} We also reproduce the
\textit{non-hysteretic} $\rho_{xx}$ peak at $\sim 3.2\unit{T}$
(single level crossing $\mu $) seen  in \cite{knobel}. We find that
the peak at $\sim 2.8$ T \cite{knobel} arises from \textit{two}
levels crossing $\mu$ and predict that it shifts \textit{upward} (as
opposed to Fig. 3 here) as $\theta $ increases because $g>0$ in
\cite{knobel} while $g<0$ in \cite{jaros}. However, we do not find
any hysteretic behavior here. Essentially, the non-integer filling
factors near the opposite-spin level crossings in \cite{knobel} (as
opposed to \cite{jaros}) result in exchange energy gains not enough
to induce phase transitions.

\begin{figure}[t]
\begin{center}
\includegraphics[width=8.0cm,keepaspectratio=true]{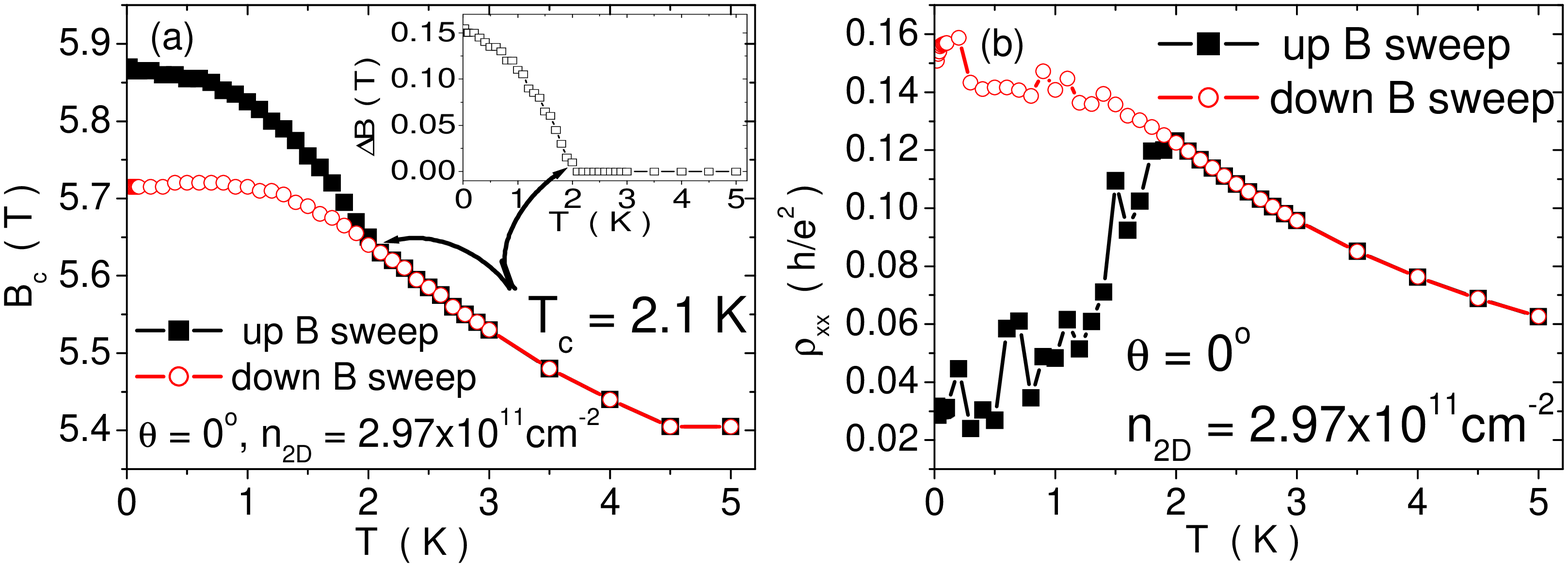}
\end{center}
\caption{Temperature dependence of the
$\protect\rho_{xx}$ spike positions $B_c$ (a) and amplitudes (b) for
up$\rightleftharpoons$down $B$ sweeps. Inset in (a): $\Delta B_c =
B_c^d-B_c^u$ versus $T$ from which we extract a critical temperature
$T_c$; for $T>T_c$ the hysteresis disappears. The asymmetric shape
of our hysteretic loops [Figs. 1(b)-(d)] is clearly manifest in
(b).} \label{fig4}
\end{figure}
\textit{Final remarks.} (i) Quenched-disorder-induced domains -- not
domains arising in metastable states \cite{jun-mac,refe} -- can lead
to transport anisotropies in 2DEGs \cite{refe1,stripes}. (ii) The
relevance of the hyperfine coupling to the hysteretic phenomena has
been recognized experimentally only in the \textit{fractional}
quantum Hall regime. We account for disorder effects only via the
broadening of the Landau levels and neglect the hyperfine coupling
and domains altogether. Our successful description of hysteretic
(quantum Hall) ferromagnetic phenomena in 2DEGs highlights the power
of DFT in a non-conventional application.

We thank R. Knobel, N. Samarth, and J. Jaroszy\'{n}ski for providing
details of the experiments and K. Capelle for helpful discussions.
JCE acknowledges enlightening discussions with J. P. Perdew, O.
Gunnarsson, M. Governale, M. R. Geller, M. E. Flatt\'e and L. J.
Sham. This work was supported by FAPESP and CNPq.


\begin{thebibliography}{99}
\bibitem{peaks-nms} We find hysteretic phenomena also in the
non-magnetic systems of \cite{hysteresis} in the integer quantum
Hall regime.
\bibitem{exp-doping} The sample in Ref. \cite{jaros} is doped on one side only.
\bibitem{self-consistency} We solve the self-consistent equations for
subsequent $B$ fields by using the solution at a particular $B$ as
the input configuration to find the solution with the field slightly
(i) increased $ B+\Delta B$ in the upsweep or (ii) decreased
$B-\Delta B$ in the downsweep. We note that two distinct solutions
arise only when $v_{xc}^{\sigma_z}\neq 0$.
\bibitem{barrier} The nature of this barrier and its collapse as the metastable
state disappears needs further investigation. However, its existence
does not require domain walls (e.g. Stoner-Wohlfarth theory of
hysteresis).
\bibitem{prange} Our $g_{ext}(\varepsilon)$ is normalized as if \textit{all}
electrons were extended (conducting). This is justified because the loss of
Hall current by the formation of localized states is exactly compensated by
the remaining extended electrons [see, R. E. Prange, Phys. Rev. B \textbf{23}, R4802 (1981)].
\end{thebibliography}
\end{document}